\documentclass[twocolumn,showpacs,showkeys,amsmath,amssymb,superscriptaddress,floatfix,nofootinbib]{revtex4}

\usepackage{graphicx}
\usepackage{bm} 
\usepackage{subfigure}
	
\def\vec#1{\mathchoice{\mbox{\boldmath$\displaystyle#1$}}
{\mbox{\boldmath$\textstyle#1$}}
{\mbox{\boldmath$\scriptstyle#1$}}
{\mbox{\boldmath$\scriptscriptstyle#1$}}}
\makeatletter

\newcommand {\rootsNN} {\mbox{$\sqrt{s_{\rm NN}}$}}
\newcommand {\kt} {\mbox{$k_{\rm T}$}~}

\newcommand {\mt} {\mbox{$m_{\rm T}$}~}
\newcommand {\mtns} {\mbox{$m_{\rm T}$}}

\newcommand {\dnde} { \mbox{$\left< dN_{ch}/d\eta \right>^{1/3}$}~}
\newcommand {\dndef} { \mbox{$\left< dN_{ch}/d\eta \right>$}}
\newcommand {\rout} {\mbox{$R_{\rm out}$}}
\newcommand {\rside} {\mbox{$R_{\rm side}$}}
\newcommand {\rlong} {\mbox{$R_{\rm long}$}}
\newcommand {\rinv} {\mbox{$R_{\rm inv}$}}

\begin{document}
 
\title{Pion, kaon, and proton femtoscopy in Pb--Pb collisions at $\sqrt{s_{\rm NN}}$=2.76
  TeV modeled in 3+1D hydrodynamics.%
}

\author{Adam Kisiel} 
\email{kisiel@if.pw.edu.pl}
\affiliation{Faculty of Physics, Warsaw University of Technology, 
ul. Koszykowa 75, PL-00662, Warsaw, Poland}

\author{Mateusz Ga\l{}a\.zyn} 
\affiliation{Faculty of Physics, Warsaw University of Technology, 
ul. Koszykowa 75, PL-00662, Warsaw, Poland}

\author{Piotr Bo\.zek} 
\affiliation{AGH University of Science and Technology, Faculty of
  Physics and Applied Computer Science, al. Mickiewicza 30, PL-30059,
  Krak\'ow, Poland}
\affiliation{The H. Niewodnicza\'nski Institute of Nuclear Physics,
  Polish Academy of Sciences, PL-31342 Krak\'ow, Poland}

 
\begin{abstract}
Femtoscopy is providing information on system size and its dynamics in
heavy-ion collisions. At ultra-relativistic energies, such as those
obtained at the LHC, significant production of pions, kaons and
protons enables femtoscopic measurements for these particles. In
particular the dependence of system size on pair momentum and particle
type is interpreted as evidence for strong collective flow. Such
phenomena are naturally modeled by hydrodynamics. We present
calculations within the 3+1D hydrodynamic model coupled to
statistical hadronization code THERMINATOR 2, corresponding to Pb--Pb
collisions at $\sqrt{s_{\rm NN}}=$2.76~TeV. We obtain femtoscopic
radii for pions, kaons, and protons, as a function of pair transverse
momentum and collision centrality. We find that an approximate
universal scaling of radii with pair transverse mass and final state
event multiplicity is observed, and discuss the consequences for the
interpretation of experimental measurements.  
\end{abstract}



\maketitle 


\section{Introduction} 
\label{sec:intro} 

The collisions of heavy-ions at ultra-relativistic energies have been
studied at the Relativistic Heavy-Ion Collider (RHIC) at Brookhaven  
National Laboratory (BNL) for Au ions at $\sqrt{s_{\rm NN}}=200$~GeV. 
In the system created in such collisions the deconfined state of
strongly-interacting matter is created, where the relevant degrees of
freedom are quarks and gluons. It behaves as a strongly
coupled liquid with small
viscosity~\cite{Adams:2005dq,Adcox:2004mh,Back:2004je,Arsene:2004fa}.
This behavior is well described by 
a variety of hydrodynamic codes, in terms of the transverse momentum
spectra and the radial and elliptic flow phenomena, with the important
addition of event-by-event fluctuations. While the behavior of the
system is well understood in the momentum sector, its description in
the space-time domain remained a challenge, and was achieved a few
years ago~\cite{Broniowski:2008vp,Pratt:2009hu}. Based on these
findings, hydrodynamic models were adapted 
to heavy-ion collisions at the highest currently available energy,
achieved at the Large Hadron Collider (LHC) at CERN, for Pb ions with
$\sqrt{s_{\rm NN}}=2.76$~TeV. 

One of the important cross-checks of the collective picture of the
heavy-ion collision is the investigation of the space-time scales of
the system, obtained via femtoscopy for pairs of identical
particles. Such analysis is performed in the Longitudinally Co-Moving
System (LCMS), where the the longitudinal direction is along the beam
axis, the outwards direction is along the pair transverse momentum and
the sidewards direction is perpendicular to the other two. Three
independent sizes of the system in these directions (usually referred
to as $R_{long}$, $R_{out}$, and $R_{side}$ respectively, collectively
called the ``femtoscopic radii'') are extracted as a function of event
centrality and average pair transverse momentum $k_{\rm T}=|\vec{p_{\rm T,1}} +
\vec{p_{\rm T,2}}|/2$. Both at RHIC and at the LHC a scaling was observed 
for these radii measured for pions, where they depended 
linearly on final state multiplicity \dnde and have a power-law
dependence on $m_{\rm T}=\sqrt{k_{T}^{2} +
  m_{\pi,K,p}^{2}}$~\cite{Adler:2001zd,Adams:2004yc,Adler:2004rq,Abelev:2006gu,Abelev:2009tp,Afanasiev:2009ii,Aamodt:2011mr}
A question arises if similar scaling is indeed observed for
hydrodynamic models. In such calculations the hydro stage is usually
followed by statistical hadronization, and subsequent resonance
propagation and decay, as well as hadronic rescattering. Even if the
scaling is observed at the end of the hydro stage, it is not obvious
if it would still be observed after the hadronic stage. This work
provides arguments to the discussion of such questions.

Hydrodynamic collectivity has a particular feature of involving all
types of particles, including pions, kaons and protons, which are all
subject to the same flow field. Therefore it is natural to expect that
the scaling of radii will extend to results for heavier particles,
such as kaons and protons. It can be shown analytically that a
power-law scaling of the form $m_{\rm T}^{1/2}$ arises in a
one-dimensional hydrodynamic expansion with negligible transverse flow
and common freeze-out criteria~\cite{Makhlin:1987gm,Sinyukov:1989xz}.  
It is not known, if such scaling will still be present in
state-of-the-art, fully three-dimensional calculations with
significant transverse flow and viscosity, 
which are necessary to realistically model heavy-ion collisions.
In addition the hadronic rescattering and resonance decays may
have significantly different influence on different particle
types. A recent realistic calculation including a hydro phase as
well as hadronic rescattering phase suggests that the scaling between
pions and kaons is broken at the LHC~\cite{Shapoval:2014wya}. It is 
important to know, if this arises already at the hydro phase or if it
is the effect of the hadronic rescattering. In the first case, the
argument 
of the \mt scaling for radii for different particles as a signature of 
collectivity should be revisited. In the second case the experimental
search for such a breaking would be an excellent probe for the length
and importance of the hadronic rescattering phase at the LHC. In this
work we perform the complete calculation for pions, kaons and protons
in a model which includes the hydrodynamic phase as well as
statistical hadronization and resonance contribution, but does not
include hadronic rescattering. We discuss the consequences for the
``\mt scaling as collectivity signature'' argument. 

Kaon and proton femtoscopy is significantly more challenging than the
corresponding measurement for pions. As a consequence it is often done
not in LCMS in three dimensions, but in a simplified way, in one
dimension and in the Pair Rest Frame (PRF). We discuss what is the
relation of the scaling of femtoscopic radii in LCMS and in PRF, and
what kind of behavior is expected for one-dimensional radii for pions,
kaons, and protons, measured as a function of pair \mtns.

The calculations presented in this work serve two purposes. Firstly,
they represent predictions from 3+1-dimensional (3+1D) hydrodynamics +
THERMINATOR 2 model for pion, kaon, and proton radii at the LHC as a
function of event multiplicity and pair \mtns. Secondly, they
calibrate an important probe of collectivity: the \mt dependence of
femtoscopic radii, and give qualitative and quantitative predictions
of how collectivity can be searched for with this measurement.

\section{3+1D Hydro and THERMINATOR 2 models}
\label{sec:models}

A combination of two models is used in this work. The collective expansion
 is modeled in the 3+1D viscous hydrodynamics. The details of the
implementation and the formalism of the model is presented
in~\cite{Bozek:2011ua}. The calculation is coupled to the statistical
hadronization and resonance propagation and decay simulation code
THERMINATOR 2 \cite{Chojnacki:2011hb}.

The viscous hydrodynamic model evolves the flow velocity $u^\mu$ and 
energy density $\epsilon$ in 3+1D, following the second order 
 Israel-Stewart 
equations \cite{Gale:2012rq}. The energy momentum tensor is composed of the ideal fluid part, the stress tensor $\pi^{\mu\nu}$ and the bulk viscosity correction $\Pi$
\begin{equation}
T^{\mu\nu}=(\epsilon+p+\Pi)u^\mu u^\nu-(p+\Pi) g^{\mu\nu}+\pi^{\mu\nu} 
\ .
\end{equation}
To obtain a realistic flow profile, important for the description of the femtoscopic radii, a hard equation of state should be used \cite{Broniowski:2008vp,Pratt:2009hu}. We use a parametrizaton of the equation of state interpolating between lattice QCD results \cite{Borsanyi:2010cj} at high temperatures and the hadron gas equation of state at low temperatures.
For central rapidity region of heavy-ion collisions at the LHC, we set
all chemical potentials to zero.

 In this work we use smooth initial conditions for the hydrodynamic evolution. Event-by-event fluctuations
 in the initial conditions generate, after hydrodynamic evolution, 
fluctuating freeze-out hypersurfaces. It has been shown that these effects have negligible influence on the azimuthal angle averaged femtoscopy analysis \cite{Bozek:2014hwa}.
The initial entropy density profile in the transverse direction is given by the Glauber model, as a combination of the density of participant 
nucleons $\rho_{part}$ and  binary collisions $\rho_{bin}$ 
$\frac{1-\alpha}{2}\rho_{part}+\alpha \rho_{bin}$ with $\alpha=0.15$.
This choice of the initial density describes fairly well the centrality dependence of the charged particle density \cite{Bozek:2012qs}.
The detailed form of the entropy density profile in the longitudinal direction and the parameters can be found in Ref. \cite{Bozek:2014hwa}.
The initial time for the hydrodynamic evolution is $0.6$~fm/c, viscosity coefficients are $\eta/s=0.08$ and $\zeta/s=0.04$,  and the 
freeze-out temperature $T_t=140$~MeV. 
The calculation is performed for seven sets of initial conditions,
corresponding to the given impact parameter $b$ values (in fm) for the
Pb--Pb collisions at the \rootsNN=2.76 TeV: 3.1, 5.7, 7.4, 8.7, 9.9,
10.9, and 11.9. They correspond, in terms of the average particle
multiplicity density \dndef, to given centrality ranges at the
LHC~\cite{Aamodt:2010cz}: 0-10\%, 10-20\%, 20-30\%, 30-40\%, 40-50\%,
50-60\%, and 60-70\%. For completeness, some calculations have been
also done for $b=2.3$~fm corresponding to the 0-5\% centrality range.

 The freeze-out hypersufaces obtained in the hydro
calculation are a direct input for the THERMINATOR 2 code, which
performs a 
hadronization at these surfaces, with particle
yields following the Cooper-Frye formula
\begin{equation}
E\frac{d^3N}{dp^3}=\int d\Sigma_\mu p^\mu f(p_\mu u^\mu) \ .
\label{eq:statem}
\end{equation}
$d\Sigma_\mu$
 is the integration element on the freeze-out hypersurface and
\begin{equation}
f=f_0+\delta f_{shear}+\delta f_{bulk} 
\end{equation}
 is the momentum distribution including nonequilibrium corrections. 
The hydrodynamic evolution generates the flow velocity at freeze-out, as well as the stress  and bulk tensors, $\pi^{\mu\nu}$ and $\Pi$, necessary to calculate nonequilibrium corrections to the equilibrium distribution functions
at freeze-out 
from shear
viscosity 
\begin{equation}
\delta f_{shear}= f_0
\left(1\pm f_0 \right) \frac{1}{2 T_f^2 (\epsilon+p)}p^\mu p^\nu \pi_{\mu\nu}
\end{equation}
and bulk viscosity
\begin{equation}
\delta f_{bulk}= C f_0
\left(1\pm f_0 \right)\left(\frac{(u^\mu p_\mu)^2}{3 u^\mu p_\mu}-c_s^2 u^\mu p_\mu 
\right) \Pi  \ ,
\label{eq:fbulk}
\end{equation}
where $c_s$ is the sound velocity and
\begin{equation}
\frac{1}{C}= \frac{1}{3}\sum_{hadrons}\int \frac{d^3 p}{(2\pi)^3}\frac{m^2}{E}f_0
\left(1\pm f_0 \right)\left(\frac{p^2}{3 E}-c_s^2 E \right) \ .
\end{equation}

The THERMINATOR 2 model equates the chemical and kinetic freeze-out,
and does not include the hadronic rescattering. Nonequilibrium terms
\eqref{eq:fbulk} introduce corrections to particle rations,  the
effective chemical freeze-out temperature is higher than $T_f$
\cite{Bozek:2012qs}.  All known resonances are used in the
hadronization process. They are subsequently allowed to propagate and
decay, in cascades if necessary. For every particle its creation point
is either located on 
the freeze-out hypersurface (so-called ``primordial'' particles) or is
associated with the point of the decay of the parent particle. This
information is crucial for femtoscopic analysis performed in this
work and is kept in the simulation.

\section{Femtoscopic formalism}
\label{sec:femtoform}

The femtoscopic correlation function is a ratio of the conditional
probability to observe two particles together, divided by the product
of probabilities to observe each of them separately. Experimentally it
is measured by dividing the distribution of relative momentum of pairs
of particles detected in the same collision (event) by an equivalent
distribution for pairs where each particle is taken from a different
collision. 
The femtoscopy technique focuses on the mutual two-particle
correlation, which can come from wave-function (anti-)symmetrization for
pairs of identical particles. In this case the measurement is
sometimes referred to as ``Hanbury-Brown Twiss (HBT) correlations''. Another source is the
Final State Interaction (FSI), that is Coulomb or strong. 
At the moment no heavy-ion collision models exist that would take the
effects of two-particle wave-function symmetrization or the FSI into
account when simulating particle production. 
The effect is usually added in an ``afterburner'' code. This procedure
is also used in this work. It requires the knowledge
of each particles' emission point and momentum, which is provided by
the THERMINATOR 2 model. We do not compare our correlation functions
with the experiment directly, the comparison is only done at the level
of the extracted femtoscopic radii. In the experimental analysis the
FSI is usually a methodological complication, while the main physics
observable is the correlation resulting from quantum statistics (QS)
(anti-)symmetrization. Therefore in this work we simplify the
``afterburner'' calculation and only take into account the QS effect,
the FSI is not taken into account. The procedure to extract the
femtoscopic radii is modified accordingly and assumes that the only
source of correlation 
is the QS. With this simplification the calculated radii can still be
compared with experimental ones, while the systematic uncertainty on
the calculation is reduced.

With such assumptions the correlation function can be expressed as: 
%
\begin{equation}
C(\vec{k^{*}}) = {{\int S(\bf{r^{*}}, \vec{k^{*}})
  |\Psi(\bf{r^{*}}, \vec{k^{*}})|^{2}} \over {\int
  S(\bf{r^{*}}, \vec{k^{*}})}} 
\label{eq:cfrompsi}
\end{equation}
where $\bf{r^{*}}={\bf x}_{1}-{\bf x}_{2}$ is a relative space-time
separation of the two particles at the moment of their
creation. $\vec{k^{*}}$ is the momentum of the first particle in the 
PRF, so it is half of the pair relative momentum in
this frame. $S$ is the source emission function and can be interpreted
as a probability to emit a given particle pair from a given set of
emission points with given momenta. 
For identical bosons (pions and kaons), the wave function must be
symmetrized and takes the form:
\begin{equation}
\Psi_{\pi,K} = 1+\cos(2\vec k^{*} r^{*})
\label{eq:psibosons}
\end{equation}
while for the unpolarized fermions (protons), it is:
\begin{equation}
\Psi_{p} = 1-\frac{1}{2} \cos(2\vec k^{*} r^{*})
\label{eq:psifermions}
\end{equation}
The calculation of the correlation function is as follows. First all
particles of a certain type (charged pion, charged kaon, proton) from
a THEMINATOR 2 event for a given centrality are combined into
pairs. A histogram $B$ is created where each pair is filled with the
weight of 1.0, at a corresponding relative momentum
$\vec{q}=2\vec{k^{*}}$. The histogram can be one-dimensional (as a
function of $|\vec{q}|$), three dimensional (as a function of three
components of $\vec{q}$ in LCMS), or a set of one-dimensional
histograms representing selected components of the spherical harmonic
decomposition of the distribution~\cite{Kisiel:2009iw}. The second
histogram $A$ is created, where the pair is inserted in the same
manner, but with the weight calculated according to
Eq.~\eqref{eq:psibosons} for pions and kaons or
Eq.~\eqref{eq:psifermions} for protons. The correlation function $C$
is calculated as $A/B$. Mathematically this procedure amounts to a
Monte-Carlo calculation of the integral given in
Eq.~\eqref{eq:cfrompsi}. The $C$ obtained in this way closely
resembles an experimental correlation function (modulo the absence of
the FSI) and a standard
experimental procedure to extract the femtoscopy radii from it can be
applied. It is also identical to the procedure used in previous
calculations based on the THERMINATOR
model~\cite{Kisiel:2006is,Kisiel:2008ws,Bozek:2011ua}.

\begin{figure}[tb]
\begin{center}
\includegraphics[angle=0,width=0.45 \textwidth]{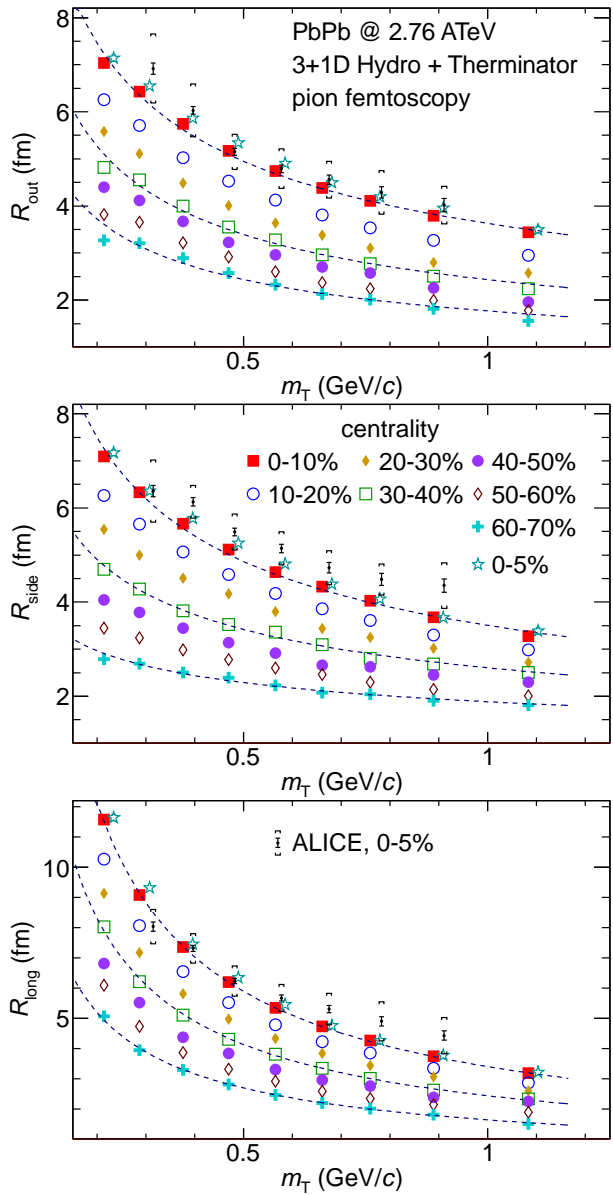}
\end{center}
\vspace{-6.5mm}
\caption{Femtoscopic radii in LCMS calculated for pions, as a function
  of pair transverse momentum and centrality. Dashed lines show
  power-law fits to selected centralities. For completeness,
  calculations for $b=2.3$~fm are compared to ALICE
  data~\cite{Aamodt:2011mr} at the corresponding centrality. The two
  data sets are slightly shifted in $x$ direction for visibility.
\label{fig:pioncentmt}}
\end{figure}

The correlation functions are then fitted to extract the femtoscopic
radii, in a procedure designed to closely resemble the experimental
one. First, the functional form of $S$ is assumed to be a
three-dimensional ellipsoid with a Gaussian density profile:
\begin{equation}
S(\vec{r}) \approx \exp \left( -\frac{r_{out}^{2}} {4R_{out}^{2}}
-\frac{r_{side}^{2}} {4R_{side}^{2}} -\frac{r_{long}^{2}}
{4R_{long}^{2}}\right ) ,
\label{eq:sgauslcms}
\end{equation}
where $r_{out,side,long}$ are components of $r^{*}$ calculated in
LCMS and $R_{out}$, $R_{side}$, and $R_{long}$ are single-particle
femtoscopic source radii. Then Eq.~\eqref{eq:cfrompsi} gives the
following fit function:
\begin{equation}
C(\vec{q}) = 1+\lambda \exp \left(-R_{out}^{2}
  q_{out}^{2}-R_{side}^{2}q_{side}^{2} -R_{long}^{2}q_{long}^{2}
\right) .
\label{eq:cfit}
\end{equation}
It can be directly fitted to the calculated correlation functions to
extract the femtoscopic radii. Because of the simplifying assumption
of not including the FSI in the calculation of the model function,
there is no need for additional factors 
accounting for them in the fitting function. The formula can be used
directly for the three-dimensional function in Cartesian
representation as well as in spherical harmonics decomposition. For
one-dimensional correlation function a simplified source assumption is
made:
\begin{equation}
S(\vec{r^{*}}) \approx \exp \left( -\frac{{r^{*}}^{2}} {4R_{\rm inv}^{2}}
\right ) ,
\label{eq:sgausprf}
\end{equation}
where the source is spherically symmetric in PRF with a single source
size $R_{\rm inv}$. This gives the one-dimensional fit function:
\begin{equation}
C(q_{\rm inv}) = 1+\lambda \exp \left(-R_{\rm inv}^{2}
  q_{\rm inv}^{2}\right) .
\label{eq:cfitprf}
\end{equation}

\section{Source sizes for pions, kaons, and protons}
\label{sec:pikpradii}

The correlation functions are calculated separately for pions, kaons,
and protons, for seven centrality ranges. The momentum dependence is
studied by calculating the correlation function separately for pairs
in the following \kt ranges (given in GeV/$c$): 0.1-0.2, 0.2-0.3,
0.3-0.4, 0.4-0.5, 0.5-0.6, 0.6-0.7, 0.7-0.8, 0.8-1.0, and 1.0-1.2 for
pions. For kaons the \kt ranges start from 0.3, for protons from
0.4, since at lower momenta the multiplicity of the heavier particles
is too limited to perform a reliable calculation. The given \kt ranges
contain the significant part of the experimental acceptance at the
LHC, in particular the acceptance of the ALICE experiment, which is
the only one with advanced particle identification capabilities
applicable in high multiplicity events on a particle-by-particle
level. The calculation is intended as a test of the \mt scaling of
radii for different particle types, therefore it is important that the
\mt ranges for pions, kaons and protons overlap significantly. 

\begin{figure}[tb]
\begin{center}
\includegraphics[angle=0,width=0.45 \textwidth]{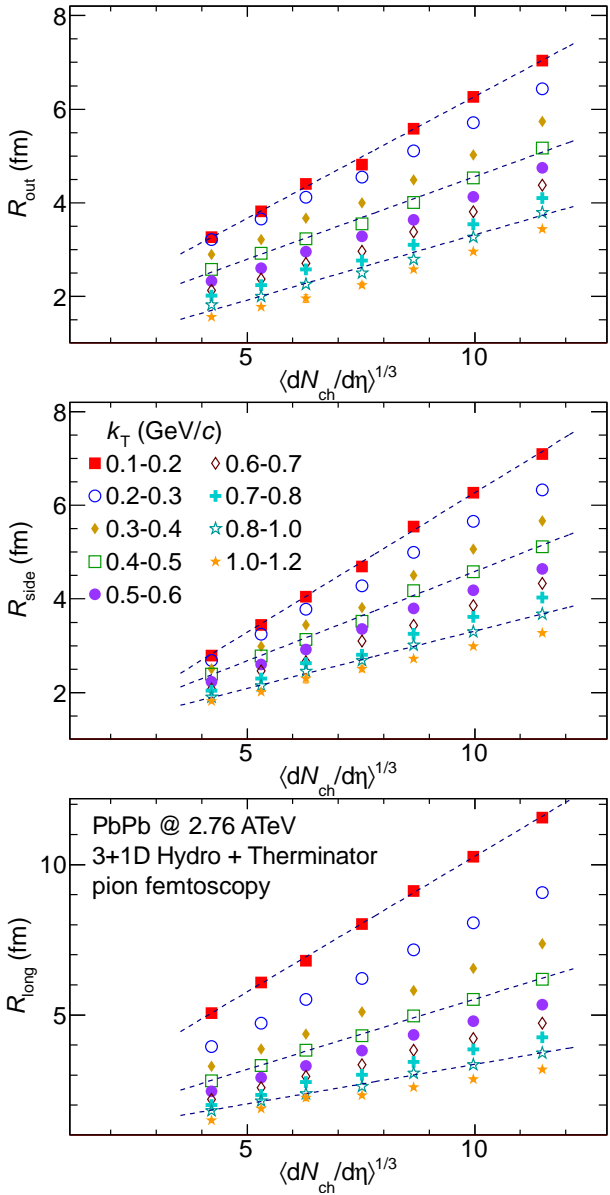}
\end{center}
\vspace{-6.5mm}
\caption{Femtoscopic radii in LCMS for pions as a function of cube
  root of the charged   particle multiplicity for several \kt
  ranges. Lines represent linear fits to selected \kt ranges.
\label{fig:piscalingmult}}
\end{figure}

In Fig.~\ref{fig:pioncentmt} the radii in LCMS for pions are shown as
a function of centrality and \mtns. The transverse size reaches 7~fm for
the lowest \mt and largest multiplicity events, the longitudinal size
reaches over 11~fm in the same range. The lowest radii observed are on
the order of 2~fm, for high \mt at the largest centrality. The
calculations for $b=2.3$~fm are in good agreement with data from
top 5\% central collisions from ALICE~\cite{Aamodt:2011mr}. The radii
universally fall with  \mtns, in all directions and for all
centralities. The lines, drawn for selected centralities, show fits
of the power-law function: 
\begin{equation}
f(m_{\rm T}) = \alpha m_{\rm T}^{\beta} ,
\label{eq:powerlawfit}
\end{equation}
where $\alpha$ and $\beta$ are free parameters. In all cases the
power-law type function fits the radii dependence on \mt well. For the
$out$ radius, the $\beta$ parameter is on the order of -0.45. For the
$side$ radius it is similar, however for the highest centrality it is
closer to zero. For the $long$ radius the slope is steeper, resulting
in the $\beta$ of -0.75, also slightly closer to zero for highest
centrality. This behavior, known as the ``lengths of homogeneity''
mechanism~\cite{Makhlin:1987gm,Akkelin:1995gh}, is a signature of the
collective flow of the system.

The radii universally grow with decreasing centrality (increasing
event multiplicity), in all directions and at all \mt. In
Fig.~\ref{fig:piscalingmult} the same radii are re-plotted as a
function of the final state multiplicity. The lines represent fits to
the selected \mt ranges, linear in \dnde. They show that the
dependence is indeed universally linear in this variable, in all
directions and at all centralities. 

The two scalings mentioned above can be generalized with a common
formula for all centralities and all \mtns, given below:
\begin{equation}
R(x, y) = \alpha y^{\beta} \left( a+ d x \right) ,
\label{eq:rxyscaling}
\end{equation}
where $y$ is \mt and $x$ is \dnde. The scaling holds to within 5\% for
$R_{out}$, $R_{side}$, and $R_{long}$, except for the highest centrality
where deviations can reach 10\% at high \mt. The parameters are:
$\alpha = 1.98$, $\beta = -0.46$, $a=0.33$, $d=0.128$ for $R_{out}$, 
$\alpha = 2.00$, $\beta = -0.44$, $a=0.29$, $d=0.131$ for $R_{side}$, and 
$\alpha = 1.97$, $\beta = -0.78$, $a=0.26$, $d=0.130$ for
$R_{long}$. $\alpha$ is similar for all radii, while $\beta$ is larger
in the $long$ direction. The \dnde scaling parameters are similar in
all directions. The scaling behavior shows that hydrodynamics produces
common collective behavior in both transverse dimensions. The flow in
the longitudinal direction has comparable features, but produces a
steeper dependence on \mtns, a consequence of a larger flow velocity.

\begin{figure}[tb]
\begin{center}
\includegraphics[angle=0,width=0.45 \textwidth]{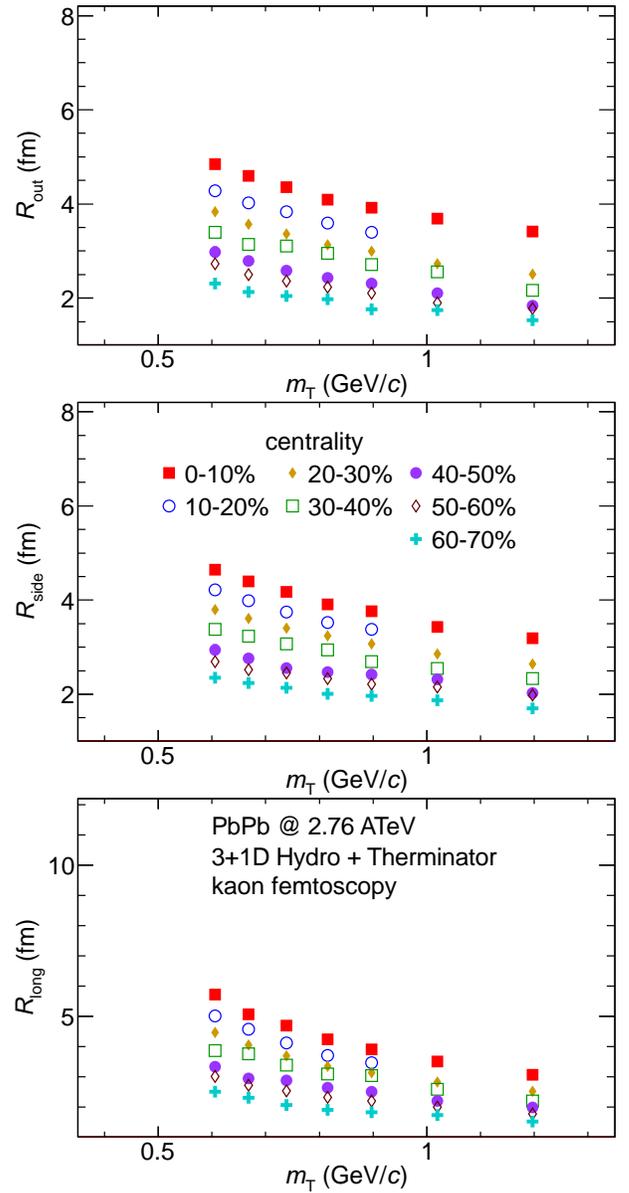}
\end{center}
\vspace{-6.5mm}
\caption{Femtoscopic radii in LCMS calculated for kaons, as a function
  of pair transverse momentum and centrality.
\label{fig:kaoncentmt}}
\end{figure}

The results for kaons are shown in Fig.~\ref{fig:kaoncentmt}. The kaon
radii also decrease with \mt and increase with \dnde. We have
performed a fit with Eq.~\eqref{eq:rxyscaling} to the kaon data and
found that the radii follow the scaling with a comparable accuracy of 
5\%. The resulting parameters were the same as for the pion case,
$\alpha$ on the order of 2.0, $a$ on the order of 0.3 and $d$ on the
order of $0.13$. Some difference was observed only for the $\beta$
exponent, which was lower for kaons, $-0.59$ in $out$, $-0.54$ for
$side$, and $-0.86$ for $long$. Taken at face value, the $\beta$
parameter difference means that there is no common scaling of radii
between pions and kaons. In reality, taking into account the fact that
experimental accuracy is seldom better than 5\%, these values indicate
that the common effective scaling between pions and kaons for radii
vs. multiplicity and pair \mt exists. We will further test with 
what accuracy such statement can be made.

\begin{figure}[tb]
\begin{center}
\includegraphics[angle=0,width=0.45 \textwidth]{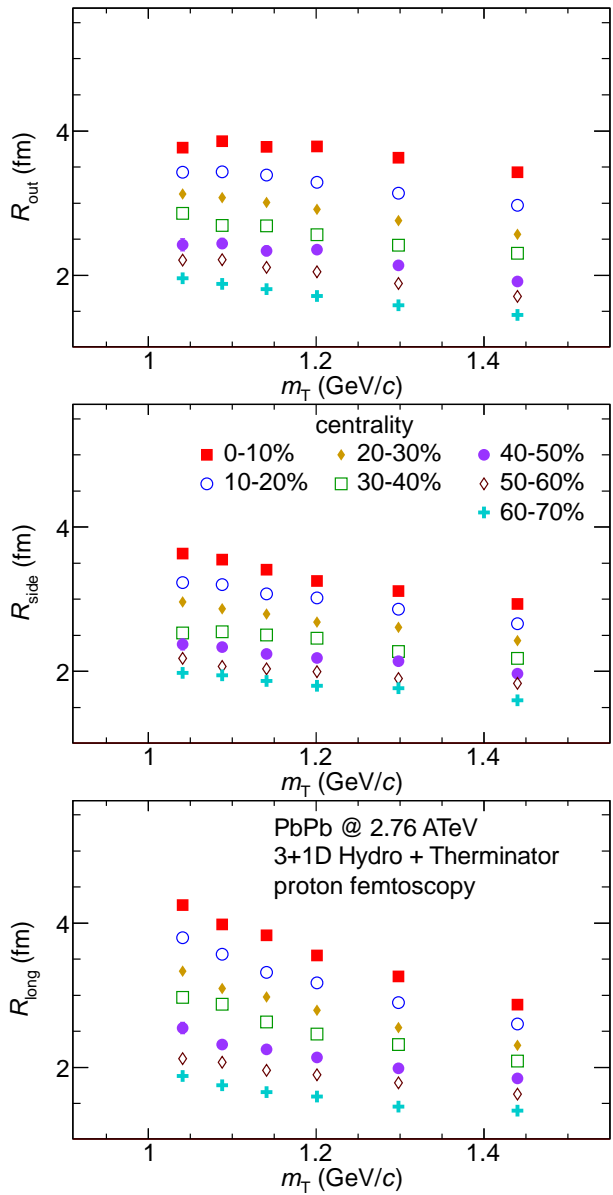}
\end{center}
\vspace{-6.5mm}
\caption{Femtoscopic radii in LCMS calculated for protons, as a function
  of pair transverse momentum and centrality.
\label{fig:protoncentmt}}
\end{figure}

Finally, the results for protons are shown in
Fig.~\ref{fig:protoncentmt}. The proton radii also decrease with \mt
and increase with \dnde. We have performed a fit with
Eq.~\eqref{eq:rxyscaling} to the proton data and found that the radii
follow the scaling with a comparable accuracy of 5\%. The resulting
parameters were the same as for pions and kaons, again the only
parameter showing difference was the $\beta$ exponent, which was
even lower for protons, $-0.58$ in $out$, $-0.61$ for $side$ and
$-1.09$ for $long$. These values are again different than for the
other particles, however close enough, so that an effective common
scaling between pions and kaons vs. multiplicity and pair \mt may
extend to protons as well.

\begin{figure}[tb]
\begin{center}
\includegraphics[angle=0,width=0.45 \textwidth]{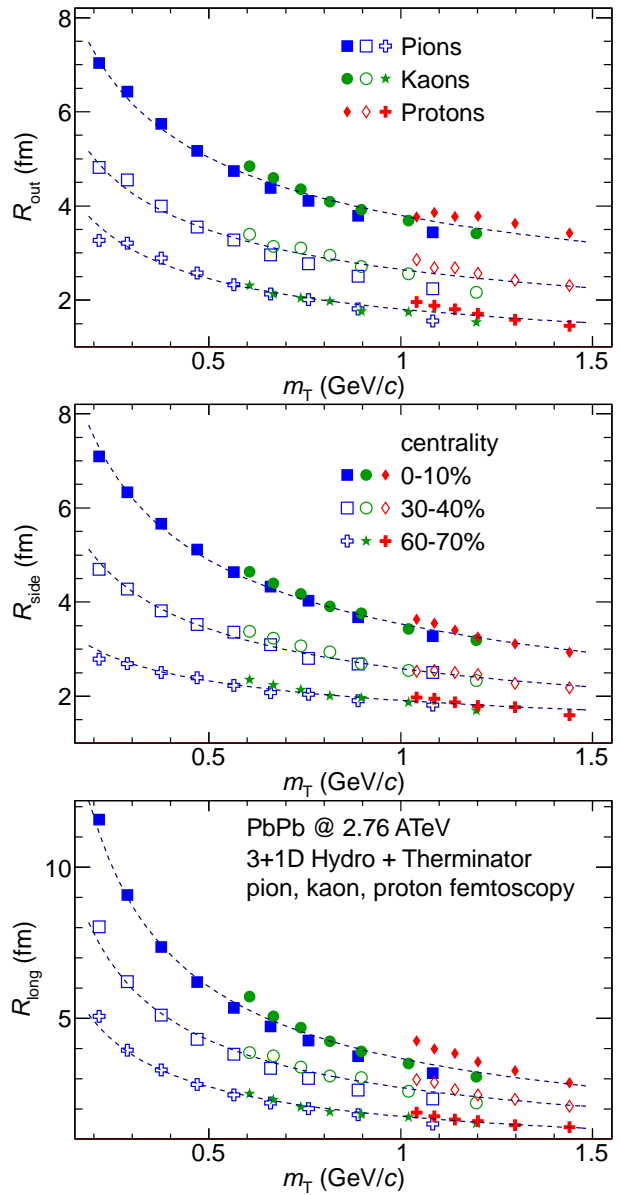}
\end{center}
\vspace{-6.5mm}
\caption{Femtoscopic radii in LCMS for pions, kaons and protons, for
  selected centralities. Lines represent power-law fits to the
  combined pion, kaon, and proton data points at a given centrality
  and direction (see text for details).
\label{fig:pikpscaling3d}}
\end{figure}

In order to test the validity of such concept, the pion, kaon, and
proton radii are plotted simultaneously as a function of \mt for
selected centralities in Fig.~\ref{fig:pikpscaling3d}. Each set of
pion, kaon and proton radii for a given direction and centrality is
then fit with a single function of the from given by
Eq.~\eqref{eq:powerlawfit}. The fits are reasonable in all cases. We
calculate the average absolute deviation of the results from the
fitted curves. For the $out$ direction they are 3\%,  5\% and 4\% for
the 0-10\%, 20-30\% and 60-70\% centrality respectively. The $\beta$
exponent is close to -0.42 for all cases. For the $side$ direction the
agreement of the fit is even better, with the average deviations of
2\%, 2\%, and 3\% respectively. The $\beta$ parameter varies more with
centrality when compared to the $out$ case, and is between 0.28 and 0.47.
In the $long$ direction the agreement of the fits improves as
the centrality increases. The average deviation is 6\%, 5\%, and 3\%
for the three centralities. The $\beta$ exponent is smaller than in
$out$ and $side$, as expected from previous calculations. It is
$-0.72$ for the lowest and $-0.64$ for the highest centrality. In
summary the plotted scalings, while not exact, are certainly realized
with the accuracy of 5\% for all directions, all centralities and the
three particle types. 

We also fitted all pion, kaon, and proton data points with a single
function given in Eq.~\eqref{eq:rxyscaling}, for $out$, $side$, and
$long$ directions. For $out$ and $side$ this effective global scaling
is obeyed at the 5\% level, with a few outliers reaching 10\%. In the
$long$ direction the scaling is obeyed to within 10\%, with a few
outliers reaching 20\%. Therefore the minimal set of global parameters
needed to approximately describe all pion, kaon, and proton data, for
all centralities and all pair \mt are, for $out$: 
$\alpha = 2.11$, $\beta = -0.40$, $a=0.32$, $d=0.128$ for $side$: 
$\alpha = 2.04$, $\beta = -0.43$, $a=0.42$, $d=0.117$, and for $long$: 
$\alpha = 2.12$, $\beta = -0.68$, $a=0.28$, $d=0.133$.

With the given quality of the fits, comparable with the experimental
uncertainties, we claim that the 3+1D Hydro + Therminator 2 simulation
predicts an effective scaling of the three-dimensional femtoscopic
radii in LCMS, common for pions, kaons and protons. The scaling has
power-law like behavior as a function of pair \mtns, with similar
exponents on the order of -0.4, in both transverse directions, while
the exponent in the longitudinal direction is smaller, on the order of
-0.7. The scaling is also linear in \dnde, with the proportionality
coefficient of 0.12 to 0.13.  This scaling gives a powerful tool for
the prediction of femtoscopic radii at any pair \mt. We also see no
reason why it should not extend to even heavier particles, such as
$\Lambda$ baryons.  

Such a precise scaling was not observed in a recent calculation from
the HKM model for pions and kaons~\cite{Shapoval:2014wya}. There the
kaon radii were predicted to be higher than the expected trend
established for pions. It was not specified whether the scaling was
broken already at the hydrodynamic stage of the model, or whether it
arisen in the hadronic rescattering phase, modeled by UrQMD. Our
study suggests that the latter scenario is true. It is therefore of
great interest to test experimentally if such scaling exists. It will
be a crucial test of the importance of hadronic rescattering phase in
heavy-ion collisions at LHC. 
 
\begin{figure}[tb]
\begin{center}
\includegraphics[angle=0,width=0.45 \textwidth]{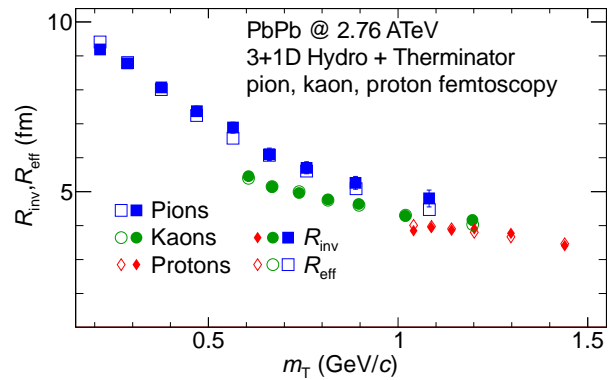}
\end{center}
\vspace{-6.5mm}
\caption{Comparison of the $R_{\rm inv}$ obtained directly from the
  fit to the one-dimensional correlation function in PRF, and a result
  $R_{\rm eff}$ of the approximate procedure to estimate $R_{\rm inv}$
  from values of $R_{\rm out}$, $R_{\rm side}$, and $R_{\rm long}$
  measured in LCMS (see text for details).
\label{fig:invapprox}}
\end{figure}

\section{Scaling in one-dimensional radii}
\label{sec:1dradii}

At LHC energies pions are most abundantly created particles, their
multiplicities per-event are large enough for a precise measurement
of pion femtoscopy in three dimensions and differentially in
centrality and \mtns. However for heavier particles, such as kaons and
protons statistics limitations arise. It is often possible to only
measure one-dimensional radius $R_{\rm inv}$ for those particles. The
measurement is then performed in the PRF. It is therefore interesting,
in view of the effective scaling of the three-dimensional radii in
LCMS shown above, to discuss similar scaling in PRF.  

The transition from LCMS to PRF is a Lorentz boost in the
direction of pair transverse momentum with velocity $\beta_{\rm T} =
p_{\rm T}/m_{\rm T}$. Therefore only the $R_{\rm out}$~radius changes, 
becoming in PRF:
\begin{equation}
R^{*}_{\rm out} = \gamma_{\rm T} R_{\rm out}.
\label{eq:routprf}
\end{equation}
where the value with asterisk is in PRF, and $\gamma_{\rm T}$ is the
Lorentz factor of the transverse boost. The measured one-dimensional
radius \rinv ~is a direction-averaged source size in PRF. According to
Eqs.~\eqref{eq:sgausprf}~and~\eqref{eq:cfitprf} it is the variance of
the Gaussian source function. One than asks how does \rinv ~depend on
the values of \rout, \rside, and \rlong? It is equivalent to the
following mathematical problem: given the three random variables $x$,
$y$, and $z$ distributed with Gaussian probability density with
variances $R_{x}$, $R_{y}$, and $R_{z}$ respectively, what is the
probability distribution of the variable $r=\sqrt{x^2+y^2+z^2}$? What
is its Gaussian width $R_{i}$? Performing a simple calculation one
finds that the
probability density of $r$ is Gaussian only in the special case of:
\begin{equation}
R_{x} = R_{y} = R_{z}.
\label{eq:equalr}
\end{equation}
In all other cases it is not Gaussian and an exact formula for
$R_{i}$ does not exist. However if the variances in three dimensions
are of the same order, the probability distribution of $r$ is
approximately Gaussian, and an ``effective'' radius can be estimated
by fitting a 
Gaussian to the Monte-Carlo simulated distribution of $r$. This is
equivalent to the ``experimentalist'' procedure of fitting the
measured correlation function with the analytical formula from
Eq.~\eqref{eq:cfitprf}, which implicitly assumes that the source is
Gaussian. 

\begin{figure}[tb]
\begin{center}
\includegraphics[angle=0,width=0.45 \textwidth]{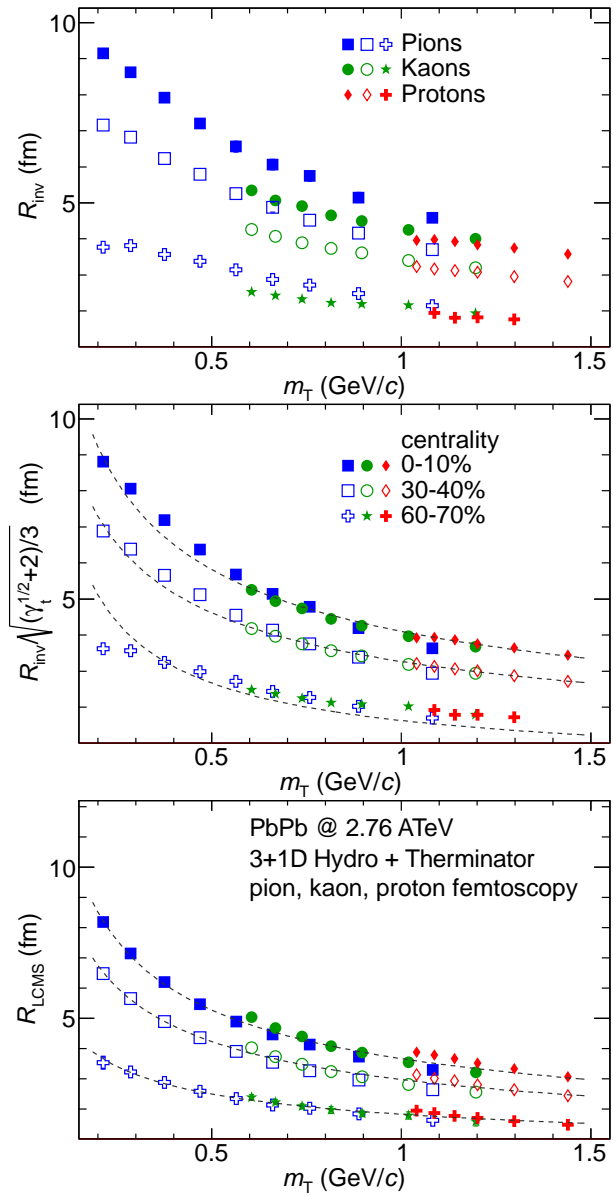}
\end{center}
\vspace{-6.5mm}
\caption{Top panel: one-dimensional femtoscopic radius $R_{\rm inv}$
  for pions, kaons, and protons calculated in the PRF for 
  selected centralities. Middle panel: $R_{\rm inv}$ for pions, kaons,
  and protons scaled with the kinematic factor, for selected
  centralities (see text for details). Lines represent power-law
  fits. Bottom panel: Averaged one-dimensional radius in LCMS $R_{\rm
    LCMS}$ 
  for pions, kaons, and protons for selected 
  centralities. Lines represent power-law fits.
\label{fig:invscaling}}
\end{figure}

We have tested this procedure for pions, kaons, and protons, the
results are shown in Fig.~\ref{fig:invapprox}. \rinv ~in the figure is
obtained from a direct fit with Eq.~\eqref{eq:cfitprf} to the
one-dimensional model correlation function in PRF. The effective
radius $R_{\rm eff}$ is calculated, 
taking the corresponding \rout, \rside, and \rlong ~from the fits
performed in LCMS (discussed in the previous section) and applying the
toy Monte-Carlo procedure described above. One can see a good
agreement of both procedures, resulting in differences between both
estimates not exceeding 3\%. In other words the \rinv value is
directly and completely determined by the values of \rout, \rside, and
\rlong in LCMS, as well as the corresponding Lorentz factor, coming
from the pair velocity.

With that introduction one can proceed to predict the scaling of \rinv
~for pions, kaons, and protons, based on the LCMS results shown in the
previous section. One immediately notices that for similar \mtns, the
$\gamma_{\rm T}$ factor for pions will be very different than the one
for kaons. A smaller, but still significant difference appears between
the factors for kaons and protons at same \mtns. As seen in
Fig.~\ref{fig:pikpscaling3d} the \rout ~for these three types of
particles is similar at same \mt, which means that $R^{*}_{out}$ for
pions, kaons, and protons will be different. That, in turn, means that
\rinv ~for pions and kaons at same \mt will differ, and so will \rinv
~for kaons and protons at same \mtns. This is indeed clearly seen in top
panel of Fig.~\ref{fig:invscaling}, where \rinv ~values for pions, kaons,
and protons show visibly different trends at any centrality. 

We summarize the discussion above by saying that a common scaling of
pion, kaon, and proton \rinv ~values {\it does not exist} (in our
model), which is a trivial kinematic consequence of the {\it
  existence} of such scaling in three dimensions in LCMS. Moreover, the
two scalings are mutually exclusive. As a consequence experimental
values of \rinv ~for pions, kaons, and protons are not good observables
in the validation of hydrodynamic collectivity predictions. For this
purpose the correct observables are radii in LCMS, measured separately
in the $out$, $side$ and $long$. 

We have argued that the violation of the scaling seen in the top panel
of Fig.~\ref{fig:invscaling} is a consequence of the large differences
in the Lorentz factor between pions, kaons, and protons at same
\mtns. If that is the case, and the \rout, \rside, and \rlong ~do scale
separately between the three particle types, then the one-dimensional
direction-averaged radius calculated in LCMS, $R_{\rm LCMS}$ should
also scale. We have performed such calculation, which is shown in the
bottom panel of Fig.~\ref{fig:invscaling}, and indeed the approximate
scaling is preserved, and still has a power-law like behavior.

We have verified that the violation of \rinv ~scaling has trivial
kinematic origin. One might then ask, if it is possible to account for
this known effect, and re-scale the measured \rinv ~for pions, kaons,
and protons in such a way that they would be a good test of the
hydrodynamic scaling. This would be experimentally much easier than
performing a full three-dimensional analysis for kaons and protons,
especially as a function of \mt. We have found that when the radii are
divided by the following scaling factor~\footnote{For the discussion
  of the origin of the precise form of the factor, please see the
  Appendix}: 
\begin{equation}
f=\sqrt{(\sqrt{\gamma_{\rm T}}+2)/3},
\label{eq:scalingfactor}
\end{equation}
they fall back on a common curve (with the accurancy of 10\%) for
pions, kaons, and protons, which 
is shown in the middle panel of Fig.~\ref{fig:invscaling}. However the
power-law behavior of the scaling, seen still for $R_{\rm LCMS}$, is
not preserved for the scaled \rinv. We have therefore given a new
``experimentalist'' recipe for the search of hydrodynamic collectivity
scaling between pions, kaons, and protons with the measurement of the
one-dimensional radius in PRF.

\section{Summary}
\label{sec:summary}

We have presented calculations of femtoscopic radii for pions, kaons,
and protons, as a function of centrality and pair \mtns. They were
performed for the 3+1D hydrodynamic model coupled to the statistical
hadronization, resonance propagation and decay code THERMINATOR
2. Hadronic rescattering was not included in the model.  The radii
were determined from the 
fits to the model correlation functions, closely following the
experimentalist's recipe. We find that the radii show two effective
scalings, which are independent of each other: 
a linear scaling in \dnde and a power-law like scaling in pair
\mt. These scalings exist separately in three dimensions in the LCMS
frame. In the two transverse directions the \mt dependence is less
steep (exponent ~-0.4) than in the longitudinal direction (exponent
~-0.7), while the \dnde scaling has 
similar slope in all directions. The scaling has common parameters for
pions, kaons, and protons (with the accuracy of 5\% to 10\%). Other
hydrodynamic calculations, which also included hadronic rescattering,
found that such scaling is violated. Therefore an experimental
verification of the existence of such scaling can serve as a probe for
the importance of the hadronic rescattering phase.

We have also discussed similar scaling for one-dimensional radii
measured in PRF. We have shown that the existence of the scaling for
the three-dimensional radii in LCMS is mutually exclusive with the
scaling for the radii in PRF between different particle types, due to
trivial kinematic reasons. We propose that a measured \rinv ~is divided
by a simple kinematic factor to recover the common effective \mt
scaling for pions, kaons, and protons. In this way an experimentally
simpler measurement of the one-dimensional radii for the three
particle types can still be used as a probe for the hydrodynamic
collectivity. 

\section*{Acknowledgment}
This work has been supported by the Polish National Science Centre
under grants No. 2011/01/B/ST2/03483, 2012/07/D/ST2/02123, and 2012/05/B/ST2/02528.

\section*{Appendix}

The form of the scaling factor for \rinv ~can be derived from the
following discussion. In the ideal case given by the
Eq.~\ref{eq:equalr}, the size \rinv ~corresponding to the variance of 
the variable $r$ is known, and equal to the other sizes. Let us
consider what will happen to the density distribution of $r$ when we
boost $x$ with some Lorentz factor $\gamma$ (corresponding to the
boost of $out$ from LCMS to PRF). It will certainly get broader,
because $x$ is now a wider distribution. A straightforward hypothesis
is that $R_{i}$ will grow as $(\gamma^{2}+2)/3$, since $R_{x}$ will
now be $\gamma R_{x}$, and the averaging between the radii in three
dimensions is done in quadrature. However Monte-Carlo simulations show
that this is not the case. As $\gamma$ grows, the $x$ distribution
becomes much wider than $y$ and $z$. As a result, the distribution of
$r$ becomes somewhat wider, but also develops long non-Gaussian
tails. Femtoscopy in this case is sensitive mostly to the width of the
distribution near the peak, and these tails will have small influence
on this width (they will have other consequences, such as lowering the
$\lambda$ parameter, but such discussion is beyond the scope of this 
paper). Therefore the $R_{i}$ grows with $\gamma$, but slower then the
naive expectation given above. We have found, through numerical
simulations, that the actual growth is best described by the factor
given by the Eq.~\eqref{eq:scalingfactor}, which has $\sqrt{\gamma}$
instead of the $\gamma^{2}$. The accuracy of such scaling is not
better than 5\%. 

\bibliography{citations}

\end{document}